\documentclass[12pt]{article}
\usepackage{amsfonts,amsthm,amscd,mathrsfs,helvet}
\usepackage{amsbsy}
\usepackage[format=hang]{caption}
\usepackage{epsfig,amssymb,amsmath,graphicx,subcaption,verbatim,hyperref,xcolor,
ulem,epstopdf,psfrag,pstool,braket,array,enumerate,multirow,wrapfig}
\usepackage{graphics,color}
\newcommand{\be}{\begin{equation}}
\newcommand{\ee}{\end{equation}}
\newcommand{\bea}{\begin{eqnarray}}
\newcommand{\eea}{\end{eqnarray}}
\newcommand{\bean}{\begin{eqnarray*}}
\newcommand{\eean}{\end{eqnarray*}}
\font\upright=cmu10 scaled\magstep1
\font\sans=cmss12
\newcommand{\ssf}{\sans}
\newcommand{\stroke}{\vrule height8pt width0.4pt depth-0.1pt}
\newcommand{\Z}{\hbox{\upright\rlap{\ssf Z}\kern 2.7pt {\ssf Z}}}

\newcommand{\C}{{\rlap{\rlap{C}\kern 3.8pt\stroke}\phantom{C}}}
\newcommand{\R}{\hbox{\upright\rlap{I}\kern 1.7pt R}}
\newcommand{\CP}{\C{\upright\rlap{I}\kern 1.5pt P}}
\newcommand{\SL}{{\rm{SL(2)}}_C}
\newcommand{\identity}{{\upright\rlap{1}\kern 2.0pt 1}}

\newcommand{\pr}{\partial}

\newcommand{\e}{{\bar e}}
\newcommand{\E}{{\bar E}}
\newcommand{\z}{{\bar z}}

\newcommand{\bphi}{{\bar \phi}}

\newcommand{\bpsi}{{\bar \psi}}
\newcommand{\bPsi}{{\bar \Psi}}
\newcommand{\mg}{\mathfrak{g}}
\newcommand{\nn}{\nonumber}
\newcommand{\Tr}{{\rm Tr}}
\begin{document}
\pagestyle{plain}

\title{\vskip -70pt
\vskip 50pt
{\bf \LARGE \bf Vortices of the Hitchin Equations}
 \vskip 30pt
}
\author{{\bf Nicholas S. Manton}
\thanks{email: N.S.Manton@damtp.cam.ac.uk} \\[15pt]
{\normalsize
{\sl Department of Applied Mathematics and Theoretical Physics,}}\\
{\normalsize {\sl University of Cambridge,}}\\
{\normalsize {\sl Wilberforce Road, Cambridge CB3 0WA, England.}}\\
}
\vskip 20pt
\date{August 2026}
\maketitle
\vskip 20pt

\begin{abstract}

The vortices that appear in Hitchin's pioneering study of Higgs bundles
are interpreted as a new version of the abelian Higgs vortices more familiar
to theoretical physicists. Their conformal covariance is made explicit.

\end{abstract}

\vskip 30pt
%PACS: 

\newpage

\section{Introduction}

In N. Hitchin's celebrated paper on Higgs bundles \cite{Hit}, vortices appear
as a special class of solutions. They satisfy an abelianised version of
what are now known as the Hitchin equations. These vortices
are defined on a compact Riemann surface and, like more general solutions
of the Hitchin equations, they are conformally invariant. Making a link to
standard abelian Higgs vortices has, as a consequence, been confusing (at least
to this author), as the standard vortices depend non-trivially on the
conformal factor of the metric of the surface where they occur; in
other words they are not conformally invariant.

There are two resolutions of this problem. One is to identify the
Hitchin vortices as abelian Higgs vortices on a surface with a
preferred metric, and the obvious choice for this is the metric
with curvature $+1$ for genus $\mg = 0$ (the round sphere), $0$ for
$\mg = 1$ (a flat torus), or $-1$ for $\mg \ge 2$ (a hyperbolic
surface). The equation for the vortex magnetic field should be adapted to the
background curvature so as to be of integrable form \cite{Five}. That
means the Popov equation on a sphere \cite{Pop}, the Jackiw--Pi
equation or Laplace equation on a torus \cite{JP1,CD}, and the more
familar abelian Higgs equation \cite{JT} or Ambj{\o}rn--Olesen
equation \cite{AO2} on a hyperbolic surface.

The second resolution is to find a modification of the standard vortex
equations so that they are explicitly conformally invariant. We show that
this can be done by replacing the usual constant term $\pm 1$ or $0$
that appears in the magnetic field equation by the local Gauss curvature
$K$. This second resolution generalises the first, and reduces to
it when the curvature is constant. For our new vortex equations,
both the Higgs field and U(1) gauge potential change in simple ways
under a conformal transformation.

The technique we use to study this problem has a large overlap with
the analysis and interpretation of vortices by Ross and Schroers
\cite{RS,Ros}, who relate integrable vortices to Cartan connections on
circle bundles over the Riemann surface. A key ingredient is a complex,
orthonormal (co)frame field $e$ on the surface, together with a real
connection 1-form $\gamma$. The structure equations relate $\gamma$ to
$e$ and its derivatives, and the Gauss curvature $K$ is the scalar
coefficient of the exterior derivative of $\gamma$. Conformal transformations
change all these quantities, and variants of these transformations
act on the Higgs and gauge fields so as to ensure the conformal
covariance of the new vortex equations.

This paper is organised as follows. After recalling the structure
equations of a Riemann surface with metric, we review the five (or six)
variants of the integrable abelian Higgs vortex equations on constant
curvature surfaces, and how to encode them in non-abelian form. We
then obtain the new, conformally-covariant vortex equations on a
surface with a general metric, determine the generalisations of the
Bradlow bound on the vortex number, find the generalised Taubes equation
for the Higgs field magnitude, and show that the curvature of the
Baptista metric associated to a vortex solution is constant. Finally,
we show that these new vortex equations are all special cases of the
Hitchin equations, so their vortex solutions are an elaboration of those
discussed by Hitchin.

\section{Frame fields on a Riemann surface}

We start by recalling the geometry of a compact Riemann surface
$\Sigma$ with metric, described in terms of a (co)frame field and connection.
Locally, we use a complex coordinate $z$ and suppose the metric is
\be
g = e^{2\rho} \, dz d\z \,,
\label{metric}
\ee
where $e^{2\rho}$ is the positive conformal factor. A choice of 1-form
(co)frame is $e = e^\rho dz$, and then $g = e\e$, where
$\e = e^\rho d\z$. The area 2-form is
\be
\omega = \frac{i}{2} e \wedge \e = e^{2\rho} \frac{i}{2} dz \wedge d\z
\,.
\label{area}
\ee
Both the metric and area 2-form are unchanged if $e$ is replaced by
$e^{i\alpha}\, e$, with $\alpha$ real and generally position-dependent.

The first structure equation (pair) is
\bea
de - i \gamma \wedge e &=& 0 \,, \nn \\
d\e + i \gamma \wedge \e &=& 0 \,.
\label{Struc1}
\eea
This introduces a real connection 1-form $\gamma$, such that the frame
is covariantly constant. The second structure equation defines the
curvature 2-form ${\cal R}$ as
\be
{\cal R} = d\gamma = K\omega \,.
\label{Struc2}
\ee
The second equality expresses ${\cal R}$ as a multiple $K$ of the
area 2-form; $K$ is then the scalar Gauss curvature of the
surface. The transformation $e \to e^{i\alpha} \, e$, accompanied
by $\gamma \to \gamma + d\alpha$, leaves the curvature fixed.

The quantities above are all locally determined by the conformal
factor of the frame, $e^\rho$. From eqs.(\ref{Struc1}) and
(\ref{Struc2}) one calculates that
\be
\gamma = i(\pr_z \rho \, dz - \pr_\z \rho \, d\z) \,, \quad K =
-e^{-2\rho} \, \nabla^2 \rho  = -\Delta\rho \,,
\ee
where $\nabla^2 = 4 \pr_z \pr_\z$ is the standard 2-dimensional
Laplacian and $\Delta =e^{-2\rho} \, \nabla^2$ is the Beltrami
Laplacian on $\Sigma$. We shall try to avoid explicit use
of $\rho$, but haven't always found this possible.

Note that $K$ is locally unconstrained, but its integral over $\Sigma$
is not, by the Gauss--Bonnet theorem. If its integral is positive,
zero or negative, respectively, then $\Sigma$ is conformally
equivalent to a sphere with $K=1$, a torus with $K=0$ or a
hyperbolic surface with $K=-1$.

\section{Vortices as flat non-abelian connections}

In this section, we review the non-abelian formulation of the equations for
abelian Higgs vortices and their variants on a Riemann surface of
constant curvature. We closely follow the analysis of Ross and Schroers --
in particlar ref.\cite{Ros}, but our conventions and notation match
those of ref.\cite{Five}. In Section 4 we will allow for conformal
transformations, which leads to the new equations for vortices on
surfaces with non-constant curvature.

We use the three real forms of the SL(2) Lie algebra, with basis
$\{ T_0, T_1, T_2 \}$ and Lie algebra brackets
\be
[T_0, T_1] = T_2 \,, \quad [T_0, T_2] = -T_1 \,, \quad [T_1, T_2] = CT_0 \,.
\ee
It is convenient to introduce $T_\pm = T_1 \pm iT_2$ and then
\be
[T_0,T_\pm] = \mp i T_\pm \,, \quad [T_+,T_-] = -2iC T_0 \,.
\label{SLalgebra}
\ee
These are the brackets for SU(1,1), SE(2) or SU(2), respectively,
when $C = -1, 0$ or $1$. We denote these three algebras
by $\SL$ with the value of $C$ indicated. We may regard $T_0, T_1,
T_2$ as real. Then, under complex conjugation,
$T_0^* = T_0$ and $T_\pm^* = T_\mp$.

The structure equations for a surface $\Sigma$
with constant curvature $K = C$ can be reformulated as the flatness of an
$\SL$ connection over $\Sigma$. It is helpful for what follows to
assume that $\Sigma$ has curvature $K = C_0$, with $C_0$
taking one of the values $-1, 0$ or $1$, independently
of the value of $C$. In terms of the frame $e$ and connection
$\gamma$ on $\Sigma$, define the real $\SL$ connection
\be
A = -\gamma T_0 + \frac{i}{2} e T_- - \frac{i}{2} \e T_+ \,.
\label{SLconnection}
\ee
Using the $\SL$ brackets we find its curvature 2-form,
$F = dA + A \wedge A$, to be
\be
F = \left( -d\gamma + \frac{i}{2} C e \wedge \e \right) T_0
+ \frac{i}{2} (de - i \gamma \wedge e) T_-
- \frac{i}{2} (d\e + i \gamma \wedge \e) T_+ \,.
\label{SLcurvature}
\ee
The last two brackets vanish by the first structure equation; by
the second structure equation, the first bracket is
$(-K + C) \omega$ (using the area 2-form formula (\ref{area})),
which equals $(-C_0 + C)\omega$. So $F=0$ and the
connection $A$ is flat if $C_0 = C$ but not otherwise.

More interestingly, we can modify the connection $A$ to
include a (complex) abelian Higgs field $\phi$ and (real) U(1)
1-form gauge potential $a$, defined on $\Sigma$. $a$ can be expressed in
terms of components as $a = a_z \, dz + a_\z \, d\z$,
with $a_\z = a_z^*$. The gauge potential has 2-form magnetic field
strength $f = da$. Requiring the flatness of this modified connection implies
that ${\phi,a}$ obey vortex equations on $\Sigma$, as we show next. We
retain independent values $-1,0$ or $1$ for $C_0$ and $C$.

On $\Sigma$, with curvature $K = C_0$, introduce
the vortex-modified frame
\be
E = \phi e \,, \quad \E = \bphi \e
\ee
and the vortex-modified connection
\be
\Gamma = \gamma + a \,.
\ee
Combine these into the modified (real) $\SL$ connection
\bea
{\hat A} &=& -\Gamma T_0 + \frac{i}{2} E T_- - \frac{i}{2} \E T_+ \nn \\
&=& -(\gamma + a) T_0 + \frac{i}{2} \phi e T_- - \frac{i}{2} \bphi \e
T_+ \,.
\label{SLvortconn}
\eea

The curvature of ${\hat A}$ is
\be
{\hat F} = \left((-C_0 + C \phi\bphi)\omega - f \right) T_0
+ \frac{i}{2} (d\phi - ia\phi) \wedge e T_-
- \frac{i}{2} (d\bphi + ia\bphi)\wedge \e T_+ \,,
\ee
which is verified as follows. From the first line of
(\ref{SLvortconn}) we obtain
\be
{\hat F} = \left( -d\Gamma + \frac{i}{2} C E \wedge \E \right) T_0
+ \frac{i}{2} (dE - i \Gamma \wedge E) T_-
- \frac{i}{2} (d\E + i \Gamma \wedge \E) T_+ \,.
\ee
Then, expanding out the second bracket gives
\bea
dE - i \Gamma \wedge E &=& d(\phi e) - i(\gamma + a) \wedge (\phi e)
\nn \\
&=& d\phi \wedge e + \phi \, de - i\phi \, \gamma \wedge e -ia \phi \wedge e
\nn \\
&=& (d\phi - ia\phi)\wedge e + \phi(de - i\gamma \wedge e) \nn \\
&=& (d\phi - ia\phi)\wedge e \,,
\label{Vorteq1}
\eea
where the last step uses the first structure equation (\ref{Struc1}).
The third bracket is handled similarly and simplifies to
$(d\bphi + ia\bphi)\wedge \e$. Finally, expanding out the first bracket gives
\bea
-d\Gamma + \frac{i}{2} C E \wedge \E
&=& -d(\gamma + a) + \frac{i}{2} C \phi\bphi \, e \wedge \e \nn \\
&=& -K\omega - da + C \phi\bphi \, \omega \nn \\
&=& (-C_0 + C \phi\bphi) \omega - f \,,
\label{Vorteq2}
\eea
where we have used the second structure equation (\ref{Struc2}).
Therefore $\hat F = 0$ provided that
\bea
(d\phi - ia\phi)\wedge e &=& 0 \,, \nn \\
(d\bphi + ia\bphi)\wedge \e &=& 0 \,, \nn \\
(-C_0 + C \phi\bphi) \omega - f &=& 0 \,.
\label{Vorteqs}
\eea

Recall that the frame $e$ is a multiple of $dz$. Therefore
these equations can be rewritten as
\bea
\pr_\z \phi - i a_\z \phi &=& 0 \,, \nn \\
\pr_z \bphi + i a_z \bphi &=& 0  \,, \nn \\
(-C_0 + C \phi\bphi) \omega - f &=& 0 \,,
\label{Vorteqs'}
\eea
the standard equations
for (variant) integrable vortices expounded in ref.\cite{Five}, and
further studied in refs.\cite{CD,Ros,Gud,MN,GR}. There are potentially nine
combinations of $C_0$ and $C$, but only five lead to non-trivial
vortex solutions. All these variants can be reduced to Liouville's equation.
The most familiar case is when $C_0 = C = -1$. This is the
abelian Higgs vortex system on a hyperbolic surface of curvature
$-1$. It was first considered on the hyperbolic plane and solved by
Witten using rational functions \cite{Wit}. A few solutions have
been constructed on compact hyperbolic surfaces with large isometry
groups \cite{MM}. Another case is the Popov vortex system on a sphere,
with $C_0 = C = 1$, whose solutions were constructed in ref.\cite{NMPop}.

\section{Conformal covariance}

It is easy to verify that the flatness of ${\hat A}$ continues to hold under
a (position-dependent) phase rotation of the frame and an independent
abelian gauge transformation of $\phi$ and $a$. However, there is a
more interesting invariance preserving the flatness of ${\hat A}$.
This is a conformal transformation that leaves ${\hat A}$ strictly unchanged.

For the (variant) vortex equations (\ref{Vorteqs'}), with
$C_0$ and $C$ fixed constants, there are no simple
transformations that preserve solutions when the background geometry
is conformally modified (changing $\omega$), but using the non-abelian
formalism here we shall find a new version of the vortex equations
where there are such simple transformations. As a consequence, the
vortex solutions of the new equations do not significantly depend on the
conformal factor of the background metric, but reduce to solutions
in the constant curvature backgrounds that we have discussed above.

To proceed, we need the formulae for the variations of the frame $e$ and
connection $\gamma$ under a conformal transformation by a
(position-dependent) factor $e^\sigma$, with $\sigma$ real. The
transformed frame is
\be
e' = e^\sigma e \,, \quad \e' = e^\sigma \e \,,
\ee
so the transformed metric and area 2-form are $g' = e^{2\sigma}g$ and
$\omega' = e^{2\sigma} \omega$, and the transformed connection is
\be
\gamma' = \gamma + i(\pr_z \sigma \, dz - \pr_\z \sigma \, d\z) \,.
\ee
These satisfy the structure equations (\ref{Struc1}) and
(\ref{Struc2}), but the curvature is modified. Recalling that $\omega
= \frac{i}{2} e^{2\rho} \, dz \wedge d\z$, we find that
\bea
d\gamma' &=& d\gamma - 2 i \, \pr_\z \pr_z \sigma \, dz \wedge d\z \nn \\
&=& K \omega - 4 e^{-2\rho} \pr_\z \pr_z \sigma \, \omega \nn \\
&=& e^{-2\sigma} (K - e^{-2\rho} \nabla^2 \sigma) \omega' \nn \\
&=& e^{-2\sigma} (K - \Delta \sigma)\omega' \,.
\eea
As $d\gamma' = K' \omega'$, the transformed curvature is
\be
K' = e^{-2\sigma} (K - \Delta \sigma) \,.
\label{Kconftrm}
\ee
In particular, if $\Sigma$ is flat, with $K=0$, then the conformal
transformation produces a surface with curvature
$ -e^{-2\sigma}\nabla^2\sigma$.

The conformal transformation of the vortex fields is arranged so that
the non-abelian connection ${\hat A}$ is unchanged. Clearly,
from formula (\ref{SLvortconn}), the transformed fields are
\be
\phi' = e^{-\sigma} \phi
\label{Higgstrans}
\ee
and
\be
a' = a -i(\pr_z \sigma \, dz - \pr_\z \sigma \, d\z) \,.
\label{gaugepottrans}
\ee
The connection is now
\be
{\hat A}' = -(\gamma' + a') T_0 + \frac{i}{2} \phi' e' T_-
- \frac{i}{2} \bphi' \e' T_+ \,,
\label{SLvortconn'}
\ee
but this equals ${\hat A}$ and therefore obviously remains flat under
these combined transformations. The transformed vortex fields satisfy
equations that can be deduced from eq.(\ref{Vorteq1}) and the
second line of eq.(\ref{Vorteq2}). These are (dropping the
primes)
\bea
\pr_\z \phi - i a_\z \phi &=& 0 \,, \nn \\
\pr_z \bphi + i a_z \bphi &=& 0  \,, \nn \\
(-K + C \phi\bphi) \omega - f &=& 0 \,,
\label{ConfVort}
\eea
where $\omega$ and $K$ are the area 2-form and curvature of the
conformally transformed surface. The first two
equations have the usual form, so the essential change is
that $K$ replaces $C_0$ in the final equation.

The conformal covariance of these new vortex equations can be
verified directly. One can make a further
conformal transformation, replacing $e$ by $e^{\chi} e$ and finding
the transformed area 2-form and curvature. Provided one then transforms
$\phi$ and $a$ analogously to (\ref{Higgstrans}) and
(\ref{gaugepottrans}) (with $\chi$ replacing $\sigma$), the
new vortex equations remain satisfied.

This conformal covariance implies that vortex solutions on a general
surface $\Sigma$ can be obtained from solutions on surfaces
with constant curvature (the integrable case). But it requires finding
the conformal transformation that changes $\Sigma$ to having constant
curvature, which is a classic problem in differential geometry.

\section{Bradlow Bound and Taubes Equation}

There is an interesting inequality satisfied by the vortex number on a
surface, due to Bradlow \cite{Bra}. For the standard (variant)
vortices, it gives bounds on the vortex number in terms of the surface
area \cite{Five}. For the new vortex equations there is an analogous, purely
topological bound.

This bound is obtained by integrating the final equation
(\ref{ConfVort}) over $\Sigma$.
As the total flux of $f$ is $2\pi$ times the vortex number $N$, and
by the Gauss--Bonnet theorem, the integral of $K\omega$ is $2\pi(2-2\mg)$
where $\mg$ is the genus of $\Sigma$, one obtains
\be
C \int_\Sigma \phi\bphi \, \omega = 2\pi (N + 2 - 2\mg) \,.
\label{Higgsint}
\ee
So, as the integral on the left-hand side is non-negative, 
\bea
N \le 2\mg - 2 \ &{\rm if}& \ C = -1 \,, \nn \\
N = 2\mg - 2 \ &{\rm if}& \ C = 0 \,, \nn \\
N \ge 2\mg - 2 \ &{\rm if}& \ C = 1 \,.
\label{Bradlowbound}
\eea
Moreover, the limiting cases $N = 2\mg - 2$ when $C = \pm 1$ are not true
vortices, as $\phi$ has to vanish everywhere. When $C = 0$, $\phi$ does
not need to vanish.

Because of the conformal invariance, it is unsurprising that these
bounds are the same as those found in ref.\cite{Five} for integrable
vortices on surfaces with constant curvature.

There is also a Taubes equation \cite{JT} for the new
vortices, an equation for the gauge-invariant Higgs field magnitude $|\phi| =
e^h$. It is obtained by using the first two equations (\ref{ConfVort})
to eliminate the gauge potential. This gives
\be
a_\z = -i\pr_\z (\log\phi) \,, \quad a_z = i\pr_z (\log\bphi) \,,
\ee
and therefore
\bea
f = da &=& (\pr_z a_\z - \pr_\z a_z) \, dz \wedge d\z \nn \\
&=& -i \pr_z \pr_\z \log (\phi\bphi) \, dz \wedge d\z \nn \\
&=& -\frac{i}{2} \nabla^2 h \, dz \wedge d\z \,.
\eea
The final equation (\ref{ConfVort}) then becomes
\be
\Delta h  - K + C e^{2h} = 0 \,.
\label{Taubes}
\ee
This is a variant of Taubes' equation for standard
vortices generalised to curved surfaces, but with the
constant term replaced by $K$. It is valid only away from the zeros
of the Higgs field (the vortex centres), because $h$ has a logarithmic
singularity at these points. The complete equation has additional
delta-function terms.

Equation (\ref{Taubes}) has a well-known interpretation in
2-dimensional geometry. If $g$ is a metric with Gauss curvature $K$,
then in order to construct a metric $\widetilde{g} = e^{2u} g$ with
prescribed Gauss curvature $\widetilde{K}$ together with
conical singularities \cite{Tro}, one needs to solve, away from
the singularities, the equation
\be
\Delta u - K + \widetilde{K} e^{2u} = 0 \,.
\ee
The logarithm of the Higgs field magnitude $u = h$ therefore solves
this equation, and constructs a metric with Gauss curvature $C$.

\section{Baptista Metric}

Baptista found an interesting reinterpretation of the standard vortex
equations in terms of the geometry of a metric on $\Sigma$ having conical
singularities at the vortex centres \cite{Bap}. If the background metric is
$g = e^{2\rho} dz d\z$ then the Baptista metric is
$g_B = |\phi|^2 g = e^{2(h+\rho)} dz d\z$. Baptista used
the Taubes equation to relate the curvatures of $g$ and $g_B$. The
conical singularity of $g_B$ at the centre of a vortex of
multiplicity $n$ has a conical excess $2\pi n$.

For the new vortex equations, Baptista's calculation simplifies.
The Baptista metric is a conformal deformation of the background metric
(with singularities), so formula (\ref{Kconftrm}) implies
that it has Gauss curvature
\be
K_B = e^{-2h}(K - \Delta h) \,.
\ee
Now, using the new Taubes equation (\ref{Taubes}) we see that this
simplifies to
\be
K_B = C \,,
\ee
a constant, whatever the curvature of the background, in agreement with the
observation at the end of Section 5. The conical singularities of $g_B$
have unchanged conical excess. 

We may add that the Baptista metric is conformally invariant for the
new vortices, because $g$ and $|\phi|^2$ transform oppositely,
according to the discussion in Section 4. The 1-form frame of the
Baptista metric is $E = \phi e$, and its connection 1-form is
$\Gamma = \gamma + a$, both of which are conformally invariant.
Therefore the curvature $K_B$ must be the same as that for
integrable vortices.  

\section{Hitchin equations} 

We finally show how these new vortices can be interpreted as solutions of
the Hitchin equations (Higgs bundles), introduced and
investigated in ref.\cite{Hit}. Vortices appear there as a special
class of solutions, in a somewhat different way than here, and
Hitchin stresses the conformal invariance of the equations, whereas here
we have seen how conformal invariance leads to the new version of the
vortex equations, where conformal transformations can be made
explicit. Nevertheless, the close relationship between these equations and
their vortex solutions is not very surprising.

The Hitchin equations are
\bea
{\bar D} \Psi &=& 0 \,, \label{Hit1} \\
F + [\Psi, \bPsi] &=& 0 \label{Hit2} \,,
\eea
defined on a compact Riemann surface $\Sigma$ with local complex
coordinate $z$. They have a gauge symmetry whose Lie algebra
we take to be $\SL$ (and not just SU(2) as in \cite{Hit}).
$\Psi$ is a 1-form, $\SL$-valued Higgs field that locally can be expressed
as $\psi \, dz$, where $\psi$ is a (complex) standard, adjoint
scalar Higgs field. $D$ is the gauge-covariant Dolbeault operator
with $\SL$ 1-form gauge potential $B$, so
\be
{\bar D} \Psi = {\bar D} \psi \, dz = (\pr_\z \psi + [B_\z , \psi])
\, d\z \wedge dz \,.
\ee
$F = dB + B \wedge B$ is the 2-form field strength of $B$. As $\Psi$
and $\bPsi$ are 1-forms, their Lie algebra bracket has a plus-sign, i.e.
\be
[\Psi, \bPsi] = (\psi \, dz) \wedge (\bpsi \, d\z)
+ (\bpsi \, d\z) \wedge (\psi \, dz)
= [\psi,\bpsi] \, dz \wedge d\z
\ee
where $[\psi,\bpsi]$ is the usual bracket. $B$ is
required to be real, and $\bPsi$ the complex conjugate of $\Psi$. It
then follows from eq.(\ref{Hit1}) that $D \bPsi = 0$.

The Hitchin equations are derived from the self-dual Yang--Mills
equations in $\R^4$ by assuming the fields are invariant under two commuting,
translational symmetries. This leads to a version of the equations on
$\R^2$ where $\Psi$ can be regarded as a scalar. But, as Hitchin
pointed out, on a general 2-dimensional surface, eq.(\ref{Hit2})
is not even coordinate invariant if $\Psi$ is a scalar, so
the formulation with $\Psi$ a 1-form $\psi \, dz$ works better. The
equations are then invariant under holomorphic coordinate
transformations, and further, are conformally invariant, as they do
not use a metric on $\Sigma$ but only the complex structure.

As Hitchin showed, eqs.(\ref{Hit1}) and (\ref{Hit2})
imply that the real, $\SL$-valued 1-form connection
\be
{\hat B} = B + \Psi + \bPsi
\ee
is flat. This is verified by calculating
\bea
{\hat F} &=& d{\hat B} + {\hat B} \wedge {\hat B} \nn \\
&=& dB + d\Psi + d\bPsi + B \wedge B + B \wedge \Psi + B \wedge \bPsi
\nn \\
&& \qquad + \Psi \wedge B + \Psi \wedge \bPsi + \bPsi \wedge B
+ \bPsi \wedge \Psi \nn \\
&=& F + [\Psi, \bPsi] + {\bar D}\Psi + D\bPsi \nn \\
&=& 0 \,,
\eea
where some terms vanish because $\Psi$ is proportional to $dz$ and
$\bPsi$ to $d\z$.

We can now identify the vortex connection 1-form ${\hat A}$, defined
in eq.(\ref{SLvortconn'}) (and dropping the primes), as a special case
of the Hitchin 1-form connection ${\hat B}$. Recall the basis
$T_0, T_+, T_-$ for $\SL$, with brackets as in (\ref{SLalgebra}) and
the reality properties $T_0^* = T_0$ and $T_\pm^* = T_\mp$. The
identification is
\be
B = -(\gamma + a) T_0 \,, \quad 
\Psi = \frac{i}{2} \phi e T_- \,, \quad
\bPsi = -\frac{i}{2} \bphi \e T_+ \,,
\ee
which is consistent with ${\hat B}$ being real. The flatness of
${\hat B}$ is equivalent to the flatness of ${\hat A}$, and from our previous
calculations we deduce that the Hitchin equations reduce to the $\SL$
vortex equations (\ref{ConfVort}) in this case.

So the new $\SL$ vortices on $\Sigma$, whose equations incorporate
the Gauss curvature $K$, are solutions of the Hitchin equations.
The conformal invariance of these equations has been made manifest
as a conformal covariance of the vortices.

Hitchin focuses on standard vortices on a surface with genus
$\mg \ge 2$, which requires $C = -1$, and finds
the first of the vortex-number bounds (\ref{Bradlowbound}).
The moduli spaces of $N$-vortices are identified as submanifolds
of the much larger moduli space of all solutions of the Hitchin equations,
and are shown to be stationary manifolds of the Morse function
\be
2i \int_\Sigma \Tr(\Psi \bPsi) \,.
\ee
This function, evaluated on $N$-vortex solutions, reduces to
\be
\int_\Sigma \phi\bphi \, \omega \,,
\ee
which is conformally invariant, and takes the value
$2\pi(2\mg - 2 - N)$, according to eq.(\ref{Higgsint}). A similar result
should hold for Popov vortices on a genus 0 surface, with $C=1$, but
it breaks down for $C=0$ as $\Tr(T_- T_+) = 0$ in that case.

The complete Hitchin moduli space is connected, so an $N$-vortex and an
$N'$-vortex on $\Sigma$ can be continuously connected there. An interpolating
curve (in particular, a gradient flow curve of the Morse function), with
parameter $s$ in an interval $I$, could possibly be interpreted as a
monopole on $\Sigma \times I$ with charge $N-N'$, because of the
magnetic flux difference on the two boundaries.

\section*{Acknowledgements}

I am grateful to Nigel Hitchin for the stimulus given to this study by
his Dirac lecture at the Centre for Mathematical Sciences, University
of Cambridge, in May 2026.

\end{document}